# Voltage-Controlled Low-Energy Switching of Nanomagnets through Ruderman-Kittel-Kasuya-Yosida Interactions for Magnetoelectric Device Applications

Bahniman Ghosh, Rik Dey, Leonard F. Register and Sanjay K. Banerjee

*Abstract*—In this letter, we consider through simulation Ruderman-Kittel-Kasuya-Yosida (RKKY) interactions between nanomagnets sitting on a conductive surface, and voltage-controlled gating thereof for low-energy switching of nanomagnets for possible memory and nonvolatile logic applications. For specificity, we consider nanomagnets with perpendicular anisotropy on a three-dimensional topological insulator. We model the possibility and dynamics of RKKY-based switching of one nanomagnet by coupling to one or more nanomagnets of set orientation. Applications for both memory and nonvolatile logic are considered, with follower, inverter and majority gate functionality shown. Sub-attojoule switching energies, far below conventional spin transfer torque (STT)-based memories and even below CMOS logic appear possible. Switching times on the order of a few nanoseconds, comparable to times for STT switching, are estimated for ferromagnetic nanomagnets.

## I. Introduction

Spin is now being considered for both memory and logic applications. Spin transfer torque (STT)-based random access memory (RAM) may provide a low power alternative for nonvolatile memory. In addition, spin-based logic may provide a nonvolatile alternative to charge-based logic [1-3], although meeting energy requirements will be much more challenging for logic. However, conventional STT-RAM relies on spin injection to a switchable free magnet through a tunnel barrier from a magnet of fixed magnetic orientation, and requiring substantial voltages and currents over the duration of the switching. Alternative approaches are being

B. Ghosh, R. Dey, L. F. Register and S. K. Banerjee are with the Microelectronics Research Center, 10100 Burnet Road, Bldg. 160, University of Texas at Austin, Austin, TX, 78758, USA (e-mail: bghosh@utexas.edu).

considered to reduce the required currents. E.g., the giant spin Hall effect in metals has been recently used to switch nanomagnets [4], which allows for greater than unity effective spin injection efficiencies—current flow parallel to the surface of the magnet rather than through it can provide multiple opportunities per electron to transfer torque to the magnet—and low-voltage operation. Topological insulators (TIs) have spin-momentum locked surface states [5-7] and, thus, may provide still better injection efficiencies within a spin-Hall geometry [8] for power efficient switching [9]. However, some of this advantage may be lost for nanoscale magnets due to the limited length of the transport path beneath the magnet, and, particularly for TIs, current shunting to a metallic magnet [10]. Voltage-aided or induced switching of nanomagnets also is being considered to reduce or eliminate the current requirement for still more power-efficient switching [11]. However, such methods rely on voltage-induced changes in the magnet's easy axis orientation and strength, requiring precise fabrication to achieve a nominal magnetic anisotropy on the boundary between vertical and in-plane orientations, and applied voltages that still would be substantial compared to those employed for CMOS logic [12].

In this work, we explore through simulation a novel mechanism for low-energy switching of the magnetic orientation of nanomagnets: voltage-controlled gating of a surface electron gas mediating Ruderman-Kittel-Kasuya-Yosida (RKKY) interaction between nanomagnets. We estimate that it may be possible to use such gated RKKY interactions to create energy-efficient memory and nonvolatile logic.

## II. Illustrative applications of gated RKKY interactions

As a prelude to our calculations, we provide illustrative examples of how such gated RKKY interactions might be used for memory and logic. Such gating of an electron gas mediating the indirect/second-order RKKY interaction between nearby nanomagnets is illustrated in Fig. 1(a). Here, for specificity we consider nanomagnets with perpendicular magnetic anisotropy (PMA) coupled through electrons within the two-dimensional (2D) surface states of three dimensional (3D) TIs with the Fermi level well above the Dirac point (via chemical or electrostatic doping). Use of PMA materials provide ease of fabrication, enhanced stability and robustness to variations in device characteristics [13, 14]. An MgO layer, deposited on top of the ferromagnets as shown could ensure the PMA. Although not inherently required for the RKKY-based switching mechanism, the spin-momentum locking of the TI surface states and associated transduction between spin current and charge current also could provide a natural mechanism for coupling to external charge-based circuitry. The gate is placed between nanomagnets, as shown in Fig. 1(a). The gate voltage controls the surface charge concentration beneath the gate to turn ON or OFF the coupling between nanomagnets in terms of raising the RKKY interaction strength above or reducing it below a threshold strength for switching. Therefore, in particular, it is not necessary to eliminate all RKKY interaction paths, which may extend around as well as below the gate, to effectively turn OFF the interaction. The gate control also could be used to vary the RKKY interaction between nanomagnets between ferromagnetic and antiferromagnetic in the ON state.

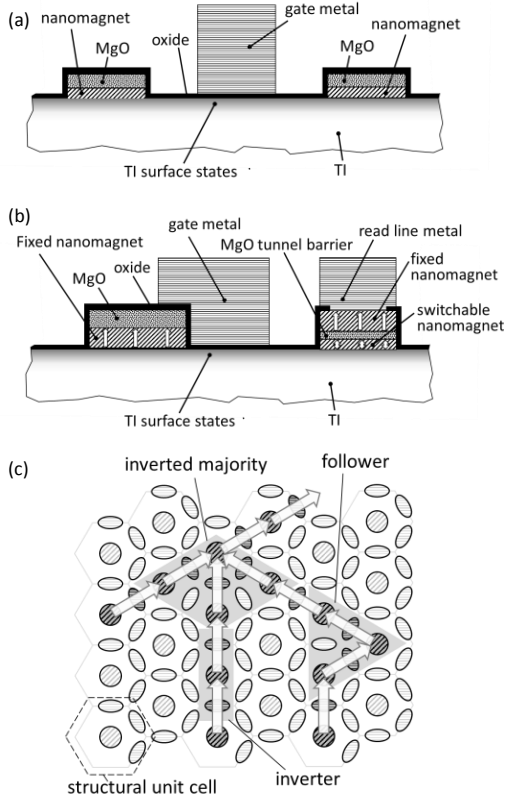

Fig. 1. (a) Side view of two nanomagnets coupled through RKKY interactions with electrons in the 2D surface states of a 3D TI, with a gate to electrostatically modulate the electron concentration between the nanomagnets to turn ON and OFF the interaction and adjust the interaction between ferromagnetic and antiferromagnetic. (b) The use of the gated RKKY to set (ON and ferromagnetic interaction), reset (ON and antiferromagnetic interaction) and store (OFF) a memory bit. (c) A gated RKKY-based reconfigurable cellular automata-like logic scheme, showing inversion, follower and inverted majority gate elements, with these functions that can overlap isolated for clarity. Normally OFF/neutral gates with antiferromagnetic coupling between all neighboring nanomagnets in the ON state are assumed in this example for simplicity. Directionality (arrows) is defined by sequential gating/clocking between adjacent ferromagnets with stabilization of "input" nanomagnets by continued coupling to one or more preceding nanomagnets.

For memory applications, RKKY coupling between a free nanomagnet could be switched between one or the other of a pair of oppositely oriented adjacent fixed magnets to set the state. Alternatively, as illustrated in Fig. 1(b),the RKKY coupling of a free nanomagnet to one adjacent fixed ferromagnet could be gated between OFF (store),ON and ferromagnetic (set), and ON and antiferromagnetic (reset).The memory state could be read in a traditional STT-RAM-like fashion (while bypassing the high-energy STT-RAM-like write).We also have taken the opportunity to illustrate that when not otherwise interfering with device requirements, the gate metal and deposited gate oxide could overlap one or both nanomagnets for a more compact layout for a given lithographic pitch, with the shown MgO layer helping to limit the overlap capacitance, the thicker the better for this purpose. Nanomagnet proximity is important not just for packing density, but because the RKKY interaction strength scales inversely with distance to the fourth power.

RKKY gating between nanomagnets also might be useable for logic within, e.g., a cellular automata-like scheme as illustrated in Fig. 1(c). Directionality of switching (arrows in Fig. 1(c) could be established by sequential gating/clocking between adjacent ferromagnets with stabilization of "input" magnets by continued coupling to one or more preceding magnets. For the shown system, we have considered a periodic array of normally OFF gates and uniformly antiferromagnetic coupling between neighboring nanomagnets in the ON state, and in what could be a reconfigurable logic scheme via the choice of which gates to switch/clock and when.

### III. Simulation methodology

To test the possibility of using RKKY switching for such applications, we consider one or more "input" nanomagnets with fixed magnetization and an "output" nanomagnet with switchable/free magnetic orientation. The ferromagnets are taken to be Fe and the topological insulator is taken to be $Bi_2Se_3$ for specificity.

The expression for the RKKY interaction energy between two magnetic point impurities on the $x$-$y$ pane surface of a topological insulator, separated in the $x$ direction but with the same position in $y$ is [15],

$$H^{RKKY} = -\frac{J^2 \varepsilon_F}{2\pi^2 \hbar^2 v_F^2 R^2}\Big[(S_1^x S_2^x + S_1^z S_2^z)\sin\left(\frac{2R\varepsilon_F}{\hbar v_F}\right) - (S_1 \times S_2)_y \cos\left(\frac{2R\varepsilon_F}{\hbar v_F}\right)\Big](1)$$

assuming a uniform carrier concentration, where: $H^{RKKY}$ is the RKKY interaction energy between the spins $S_1$ and $S_2$ of the magnetic impurities coupled through RKKY interaction, where the $x$, $y$, and $z$ superscripts and subscripts indicate the directional components thereof; $\varepsilon_F$ is the Fermi energy of the TI measured with respect to the Dirac point; $v_F$ is the Fermi velocity of the electrons, an energy independent $6.2 \times 10^5$ m/s in $Bi_2Se_3$; $R$ is the distance between the magnets on the surface of the TI; $J$ is the exchange interaction strength between $s$ orbitals in the TI and the $d$ orbitals in the FM, which is taken to be 0.13eV-nm² [15]; and $\hbar$ is the reduced Planck's constant, $1.054 \times 10^{-34}$ J-s. Weaker values of $J$, in the range 0.02eV-nm² to 0.13eV-nm², can be compensated for by smaller separations $R$ and adjustments to the Fermi level. The directional components of the magnetic field acting on the magnetic moment of Impurity 2 due to the RKKY interaction with Impurity 1 are

$$H_2^x = -\frac{J^2 \varepsilon_F}{2\gamma \pi^2 \hbar^3 v_F^2 R^2}\left[S_1^x \sin\left(\frac{2R\varepsilon_F}{\hbar v_F}\right) + S_1^z \cos\left(\frac{2R\varepsilon_F}{\hbar v_F}\right)\right] \quad (2)$$

$$H_2^y = 0 \quad (3)$$

$$H_2^z = -\frac{J^2 \varepsilon_F}{2\gamma \pi^2 \hbar^3 v_F^2 R^2}\left[S_1^z \sin\left(\frac{2R\varepsilon_F}{\hbar v_F}\right) - S_1^x \cos\left(\frac{2R\varepsilon_F}{\hbar v_F}\right)\right] \quad (4)$$

where $\gamma$ denotes the gyromagnetic ratio, $1.76 \times 10^{11}$ rad/s-T.

To simulate the magnetization dynamics, the nanomagnets are divided into cubic unit cells 1 nm on edge. As in [16], these unit cells are treated as individual impurities and a net magnetic interaction between the adjacent nanomagnets is then obtained by superposition from the individual pair-wise contributions between the unit cells of both magnets using Eqs. (2)-(4). The net magnetic field then is used to study the magnetization dynamics through the Landau-Lifshitz-Gilbert (LLG) equation [17] using the macrospin approximation,

$$\frac{d\hat{\mathbf{m}}}{dt} = -\gamma(\hat{\mathbf{m}} \times \mathbf{H}_{\text{eff}}) + \alpha\left(\hat{\mathbf{m}} \times \frac{d\hat{\mathbf{m}}}{dt}\right) \quad (5)$$

Here $\hat{\mathbf{m}}$ denotes the direction of magnetization of the output nanomagnet, and $\alpha$ is the Gilbert damping constant, taken to be 0.01, in this work. $\mathbf{H}_{\text{eff}}$ is the thus calculated net RKKY magnetic field on the output nanomagnet due to input nanomagnet(s), plus the internal demagnetization field

$$H_{\text{demag}}^x = \mu_0 M_s N_{xx} \hat{\mathbf{m}}_x, \quad (6)$$

$$H_{\text{demag}}^y = \mu_0 M_s N_{yy} \hat{\mathbf{m}}_y, \quad (7)$$

$$H_{\text{demag}}^z = \mu_0 M_s N_{zz} \hat{\mathbf{m}}_z, \quad (8)$$

plus the (interface induced) PMA field

$$\mathbf{H}_{\text{PMA}} = 2K_u \hat{\mathbf{m}}_z / M_s \quad (9)$$

where

$$K_u = K_s / t \, . \quad (10)$$

Here $\mu_0(4\pi \times 10^{-7}$ N/A²) denotes the permeability; $M_s$ is the saturation magnetization of Fe ($1.1 \times 10^6$ A/m); $N_{xx} = N_{yy}$ and

$N_{zz}$ denote the demagnetization factors [18], which are taken to be 0.07 and 0.85, respectively; $K_s$ is the PMA constant of the ferromagnet, which is taken to be $1.2 \times 10^3$ J/m² [19]; and $t$ denotes the thickness of the ferromagnet, 1 nm again.

The energy required for switching $E_s$ that required for charging and discharging the gate capacitors. Here we take that to be

$$E_s = C_g V_g^2 \quad (11)$$

as for CMOS, where $V_g$ is the gate voltage and $C_g$ is the total gate capacitance. Ignoring parasitic capacitances, we approximate the latter as the series combination of oxide capacitance and TI quantum capacitance. The TI is taken to be sufficiently thick that the capacitive coupling through the substrate can be neglected. Setting both the OFF-state Fermi level relative to the Dirac point $\varepsilon_F$ and the associated $V_g$ to zero assuming normally OFF gates (the latter via work function engineering, e.g.), the relationship between $V_g$ and $\varepsilon_F$ obtained in analogy with that in [20], is

$$\left(qV_g - \varepsilon_F\right)C_g - \frac{C_q}{2}\varepsilon_F = 0 \quad (12)$$

where $q$ denotes the charge of the electron, $1.602 \times 10^{-19}$ C. $C_q$ denotes the quantum capacitance of the surface states of TI,

$$C_q = \frac{2\varepsilon_F}{\pi(\hbar v_F/q)^2} \quad (13)$$

Solving Eqs. (13) and (14) for $V_g$ gives,

$$V_g = [1 + C_q/(2C_g)](\varepsilon_F/q). \quad (14)$$

Finally, absent an applied torque, the lifetime $\tau$ for the magnetic orientation of these macrospin states can be approximated by [21],

$$\tau = \tau_0 e^{\Delta/k_B T} \quad (15)$$

where $\tau_0 = 1$ ns, $k_B$ is Boltzmann's constant, and

$$\Delta = V[K_u - 0.5\,\mu_0 M_s^2(N_{zz} - N_{xx})] \quad (16)$$

is the thermal stability factor, where $V$ is the volume of the nanomagnet.

## IV. Results and Discussion

While nanoscales intrinsically are required, various combinations of parameters are reasonable. For the following switching simulations all of the nanomagnets are taken to be of the same square shape 15nm on an edge and of 1nm thickness for simplicity. The center-to-center distance between nanomagnets is taken to be 50nm. With the parameters assumed here, from Eqs. (15) and (16), $\tau$ is approximately 1.5 years for these macrospin nanomagnets.

As an initial example, we consider switching via RKKY coupling between a single fixed input nanomagnet and an output nanomagnet, as shown in Fig. 2(a).The RKKY coupling between the nanomagnets is taken to be initially OFF with, e.g., the Fermi level $\varepsilon_F$ close to zero for the TI surface states under the gate. At times beyond zero, it is assumed that the doping and gate voltage is such that there is a spatially uniform $\varepsilon_F$ for consistency with the assumptions underlying Eqs. (1)-(4). A spatially uniform $\varepsilon_F$ of 0.1eV in the ON state produces a predominantly ferromagnetic coupling between the nanomagnets. The RKKY interaction then triggers the switching of the output nanomagnet, as shown in Fig. 3(a). Under otherwise identical conditions, a spatially uniform $\varepsilon_F$ of 0.23eV results in a predominantly antiferromagnetic interaction in the ON state for the same inter-magnet spacing. The resulting switching dynamics are shown in Fig. 3(b). In either case, the apparent switching time is about 1 to 2ns. However, in practice longer periods would be required to ensure reliability by addressing the tail of the switching distribution much as for conventional STT-RAM. Such switching could be used for setting and resetting for memory as per the example of Fig. 1(b), or to create a logical follower and inverter, respectively.

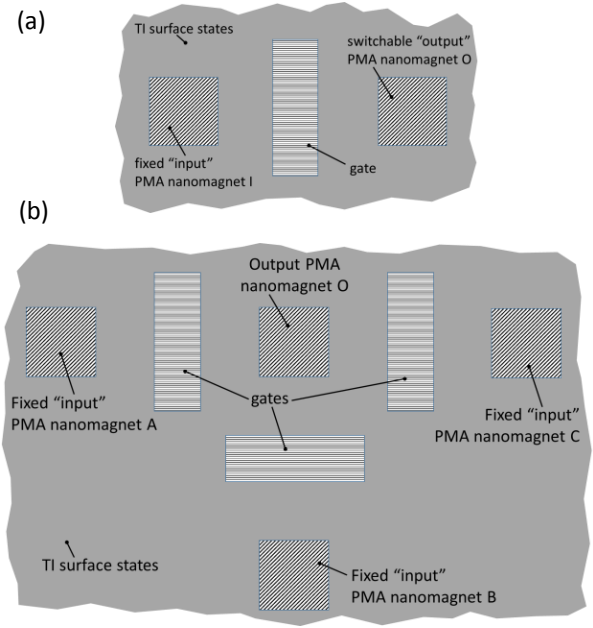

Fig. 2. Top view of simulated structures (a) for ferromagnetic (follower) and antiferromagnetic (inverter) RKKY induced switching of an output macrospin PMA magnetic O due to a fixed input macrospin PMA magnet I (results shown in Fig. 3), and (b) for RKKY induced switching in a majority gate with an output macrospin PMA magnetic O and fixed input macrospin PMA magnets, A, B, and C for ferromagnetic coupling (results shown in Fig. 4) (or inverted majority gate for antiferromagnetic coupling).

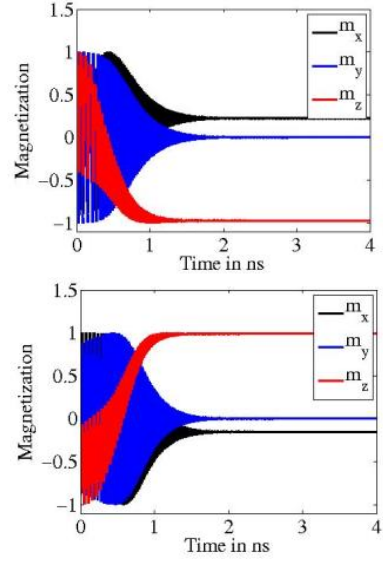

Fig. 3. Output (O) magnetization dynamics of the system of Fig. 2(a) for (a) a follower with an initially normalized output magnetization vector $\mathbf{O} = 0m_x + 0m_y + 1m_z \equiv (0,0,1)$ and input (I) vector $\mathbf{I} = (0,0,-1)$ and (b) an inverter with an initially normalized output magnetization vector $\mathbf{O} = (0,0,-1)$ and input vector $\mathbf{I} = (0,0,-1)$. Flipping the initial output and input signs produces symmetrical results in terms of the output magnetizations.

Adding a majority gate (or inverted majority gate as in Fig 1(c)) to an inverter is sufficient to achieve all Boolean logical operations. Using the same ferromagnetic RKKY inter-magnet coupling conditions as for the follower within the gate configuration of Fig. 2(b) results in the majority gate switching behavior illustrated in Fig. 4. The apparent switching time is in the 1 to 2 ns range again.

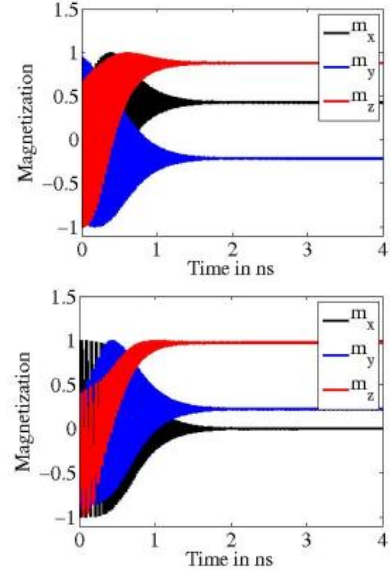

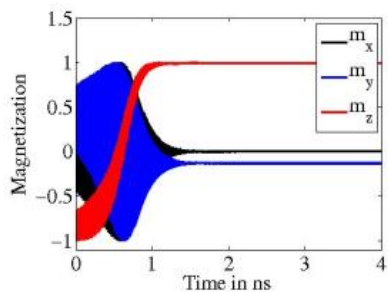

Fig. 4.Output magnetization dynamics of the majority gate output of Fig. 2(b) with an initially normalized output magnetization vector $\mathbf{O} = 0m_x + 0m_y + 1m_z$ ≡ (0,0,1) and illustrative input vectors $\mathbf{A}$, $\mathbf{B}$ and $\mathbf{C}$ of (a) (0,0,1), (0,0,1) and (0,0, −1) (b) (0,0, −1), (0,0,1) and (0,0,1), and (c) (0,0,1), (0,0,1) and (0,0,1). Flipping the initial output and inputs signs produces symmetrical results in terms of the output magnetization.

For estimating switching energies from Eqs. (11)-(14), the gate stack is taken to have an effective oxide thickness (EOT) of 0.7nm and a gate area of approximately 10nm by 30nm as depicted in Fig. 2. For the ferromagnetic and antiferromagnetic interactions, respectively, mediated by uniform electron concentrations as simulated, the $V_g$ corresponding to $\varepsilon_F$ values of 0.1eV and 0.23eV are 0.16V and 0.55V, respectively. The calculated energies for RKKY-based switching $E_s$ of the follower, inverter, majority gate, and (simulations not shown) inverted majority gate would then be 0.20 aJ, 3.5aJ 0.6aJ, and 10.5 aJ, respectively. Moreover, in principle, allowing for non-uniform carrier concentrations with a reduced electron concentration beneath the gate compensated for by somewhat higher concentration elsewhere may allow antiferromagnetic/inverter coupling with ferromagnetic/follower-like gate voltages and switching energies. However, non-uniform carrier concentrations are beyond our current switching simulation capabilities. In any case, these energies are much less than expected for current-induced switching of conventional STT-RAM memory bits, which are estimated to remain above 0.1 pJ for the foreseeable future [22], or for switching easy-plane magnets on the surface of topological insulator, estimated to be on the scale of 10-100fJ[10]. Moreover, these switching energies are comparable or smaller than perhaps even out-of-the-roadmap CMOS logic gates because of the lower gate voltages and more compact logic gates, while still allowing for nonvolatile operation. Moreover, with no "source-to-drain" voltage and normally OFF gates, there would be no significant quiescent power consumption. However, clocking remains slow compared to CMOS for ferromagnetic nanomagnet based systems, and sequential clocking is required to propagate signals over distance within a cellular-automata-like scheme.

## V. Conclusion

In this work, we proposed and explored the possibility of voltage-controlled switching of nanomagnets through RKKY inter-magnet coupling mediated by gateable electron surface charge layers, and it's possible application to low-power memory and nonvolatile logic. For specificity, we considered PMA ferromagnetic nanomagnets on the surface of a topological insulator, and exhibited switching due to both ferromagnetic and antiferromagnetic RKKY inter-magnet coupling. Switching energies were on the scale of or less than an attojoule, well below that for most nonvolatile memory concepts and even small on the scale of CMOS logic. We found switching times of the order of a few nanoseconds, which is reasonable for memory. However, even prototype experimental devices to test these concepts will have to be fabricated on the nanoscale.

## Acknowledgement

This work was supported by the NRI SWAN center, and the NSF NASCENT ERC.